\documentclass{article}
\usepackage{graphicx}
\usepackage{amsmath}
\usepackage{doublespace}
\usepackage{geometry}
\usepackage{amsfonts}
\usepackage{amssymb}

\begin{document}

\begin{center}
\textbf{Collisional Dark Matter and the Origin of Massive Black Holes}

Jeremiah P. Ostriker

\bigskip Princeton University, Princeton NJ 08544 USA
\end{center}

\qquad The nature of cosmological dark matter remains mysterious. \ Recently,
Spergel and Steinhardt [1] have revived suggestions [2, 3] that dark matter
may be strongly self-interacting,\textbf{\ }\textit{i.e.}, collisional, to
make galactic halos less dense. We show here an important by-product: \ for
reasonable values of particle mass and collisional cross-section, galaxy cores
would quite naturally grow, within them, massive black holes\textbf{\ }%
10$^{6}$\textbf{-}10$^{9}$M$_{\odot}$, having the scaling observed by
Magorrian et al. [4], $\mathbf{M}_{BH}$\textbf{\ }$\propto$\textbf{\ }%
$\mathbf{V}_{c,gal}^{4.5}$. \ 

The physical picture is quite simple. \ Assuming a normal power spectrum of
perturbations, dark matter halos begin to form in earnest in the redshift
range $Z=30\rightarrow20$, with star formation commencing in a significant way
during the interval $Z=20\rightarrow10$ (Ostriker and Gnedin [5], Haiman, Rees
and Loeb [6]). \ The massive stars formed at these early epochs (Abel et al.
[7]) will have several dramatic effects on subsequent cosmic evolution: \ they
emit UV radiation copiously, which begins to reheat and reionize the
intergalactic medium; they ultimately explode, contaminating their environment
with the first heavy elements; and, most importantly for our present purposes,
their cores implode to leave black holes (Arnett [8]), having masses $\sim$1/4
of the original stellar mass. \ The existence of a population III of high mass
stars which would have left black hole remnants is attested to by the non-zero
$(Z/Z_{\odot}\approx10^{-3})$ floor to the metallicity distribution observed
ubiquitously (9) indicating very early and widespread contamination of the
universe by ejecta from high mass supernovae. \ How such stellar mass black
holes grow to the supermassive size ($10^{6}\lesssim M_{BH}/M_{\odot}<10^{9}$)
seen in galactic nuclei [4] remains unknown, although rather general
considerations (Rees [10]) indicate the plausibility of their formation.
\ Furthermore, the quantitative scaling observed [4] $M_{BH}\propto
M_{gal}\propto V_{c}^{4.0\rightarrow4.5}$ (where $V_{c}$ is the galactic
rotation velocity) is an outstanding mystery. \ 

The 25 $M_{\odot}$ black hole remnant from a 100 $M_{\odot}$ ($H,H_{e}$) star
will immediately begin to accrete collisional dark matter in the core of the
dark halo within which it forms. \ To frame the discussion, consider the
simplest case first. \ At high densities, the dark matter will behave like an
adiabatic gas and accrete as per the classic work of Bondi [11]: \
\begin{equation}
\dot{M}_{BH}=\dot{M}_{ac}=4\pi r_{A}^{2}C_{A}\rho_{A}\;with\;r_{A}\equiv
GM_{BH}/C_{A}^{2},\tag*{(1)}%
\end{equation}
where ($\rho_{A},C_{A}$) are the density and sound speed in the ambient dark
matter fluid.

Accretion is treated in the subsequent discussion as if it were
quasi-spherical, quite different from the normal treatment of accretion from a
rotationally flattened disc. \ The reason for this is that for normal baryonic
matter, electromagnetic radiative losses are usually efficient enough so that
energy loss dominates over angular momentum loss in environments of accretion,
which leaves the contracting matter in a disc whose further evolution is
limited by the rate at which viscosity, feeding from the quasi-Keplerian
differential rotation, can transport energy and angular momentum outwards
[12]. \ But in the non-radiative case, both energy and angular momentum can be
transported only by collisions (gravitational or physical). \ Then, as shown
by Goodman [13] and others, both transport processes occur on approximately
the collisional time scale (\textit{i.e.,} the viscous and conductive time
scales are comparable), with angular momentum transported outwards rapidly
enough so that the central regions remain quasi-spherical. \ 

Stone \textit{et al.} [14] have investigated a similar situation of
non-radiative accretion in the limit of significant rotation, low viscosity
and negligible conduction. \ They convincingly find both convection and
outflow to be important, but it is not likely that their solution applies to
the high viscosity, high conductivity, low rotation case considered here. \ In
any case, most of the accretion occurs in the second, optically thin regime
which we will address shortly.

We thus return to the simple illustrative example of Bondi - accretion in a
quasi-spherical dark matter halo. \ The core region of which this matter is
composed is likely isothermal in its prior structure due to both collisional
and violent (Lynden-Bell [15]) relaxation and thus has a profile: \
\begin{equation}
\rho_{A\left(  r\right)  }=C_{A}^{2}/\left(  2\pi Gr^{2}\right)  .\tag*{(2)}%
\end{equation}
(We consider a more general profile subsequently.)

Integrating equations (1) and (2), one finds that the central massive black
hole grows at the speed of sound to reach, in time $t$, a mass $GM_{BH}\left(
t\right)  =2C_{A}^{3}t,$ where we note in passing that it will vacuum up all
the local baryonic components as well as the core dark matter.

This phase of rapid growth can only persist until the accretion radius has
grown to the point far enough from the center whereby the dark matter mean
free path approaches the accretion radius. \ Then a transition will occur to
slower, diffusively limited growth. \ At this time, t$_{1}$, mean free path
$\lambda=m_{p}/(\rho_{t_{1}}\sigma)=r_{A}=GM_{BH,t_{1}}/C_{A}^{2}$, or
\begin{equation}
GM_{BH,t_{1}}=\frac{\sigma_{p}}{2\pi}\frac{C_{A}^{4}}{Gm_{p}},\tag*{(3)}%
\end{equation}
where ($\sigma_{p},m_{p}$) are the dark matter self-interaction scattering
cross-section and particle mass respectively. \ At this time, the core radius
is r$_{t_{1}}=\sigma_{p}C_{A}/(2\pi Gm_{p}).$ \ An alternate way of writing
equation (3) in terms of dimensionless numbers is instructive:
\begin{equation}
\frac{M_{BH,t_{1}}}{m_{e}}=\frac{1}{2}\left(  \frac{e^{2}}{Gm_{e}^{2}}\right)
^{2}\left(  \frac{m_{e}}{m_{p}}\right)  \left(  \frac{\sigma_{p}}{\sigma_{T}%
}\right)  \left(  \frac{C_{A}}{c}\right)  ^{4},\tag*{(4)}%
\end{equation}
where the first quantity in parenthesis is the classic ''big number''
representing the ratio of electromagnetic to gravitational forces
($\sim10^{42}),\sigma_{T}$\ is the electromagnetic Thompson cross-section and
$c$ is the speed of light. \ To translate this into physical units, we take
$m_{p}=100G_{e}V,\sigma/\sigma_{T}=10^{-2}$ and $C_{A}=100km/s$, then
$M_{BH}=3\times10^{3}M_{\odot}$, an interesting value.

The next phase of slower growth has been treated by several authors. \ A cusp
forms about the black hole approximately described (for gravitational
interactions) by the classic Bahcall-Wolf [16] solution, as modified by loss
cone effects (Ostriker and Tremaine [17]) in the inner parts. \ The rate of
accretion is determined by the rate at which particles are scattered into the
loss cone, which begins at the Bondi accretion radius. \ Thus, it is initially
the Bondi accretion rate multiplied by the probability of strong scattering
for a particle orbiting at the Bondi radius: \ $P=\sigma C_{A}^{4}/\left(
Gm_{p}GM_{BH}\right)  $, to give an accretion rate $G\dot{M}_{BH}=2\sigma
C_{A}^{7}/(GM_{BH})(Gm_{p})$. \ This produces a mass growing, after the
transition, as the square root of the time, giving for the present time
$t_{H}$. \ $GM_{BH}=\sqrt{4\sigma C_{A}^{7}t_{H}/\left(  Gm_{p}\right)  ,} $
which gives $M_{BH}=4.1\times10^{8}$ for the previously quoted parameters, and
$t_{H}=10^{10}$ yrs. \ This produces a somewhat too large value for $M_{BH}$,
but one with approximately the correct scaling on $C_{A}$. \ In fact, this
solution cannot be extrapolated to late times or to very small values of
$\sigma/m_{p}$, because at some point accretion onto the black hole will be
limited by the mean free path in the vicinity of the Schwarzschild radius,
after which accretion only occurs within a loss cone. \ Then the estimate is
reduced by a factor of $(C_{A}/c)$ to give (Ostriker and Tremaine [17])
\begin{equation}
GM_{BH}=\sqrt{IC_{A}^{9}t\left(  \sigma/Gm_{p}\right)  /c^{2}}%
\ \ \ \ \ ,\tag*{(5)}%
\end{equation}
where the numerical constant $I$ is approximately $I=[\frac{1}{3}+2\ln
(c/C_{A})]/2\pi\approx2.60.$

Since the fundamental particle physics is quite uncertain in any case, we can
best parameterize our ignorance by defining $\eta\equiv\sigma/Gm_{p}$ and,
noting that the requirement of Spergel and Steinhardt [1] that the halo be
optically thick to collisions at a radius $r_{1}$ is equivalent to
$1\equiv4r_{1}n_{1}\sigma=2C_{A}^{2}\sigma/\pi Gm_{p}r_{1}$, or $\eta=0.5\pi
r_{1}/C_{A}^{2}$ (it is trivial to generalize these definitions to the likely
case that the cross-section is velocity dependent), we find $GM_{BH}%
\doteqdot2.0\sqrt{C_{A}^{7}t_{H}r_{1}/c^{2}}$ or $M_{BH}/M_{\odot}%
=4.9\times10^{7}C_{100}^{7/2}{}t_{H,10}^{1/2}{}r_{1kpc}^{1/2}.$ \ The range
for $r_{1}$ quoted by Spergel and Steinhardt is approximately $0.1kpcC_{100}%
^{2}<r_{1}<0.1MpcC_{100}^{2}$, or $0.45cm^{2}/g<$ $\sigma/m_{p}<450cm^{2}/g$.
\ Then, substitution into equation (5) gives $M_{BH}=7.06\times10^{6}%
(\sigma/m_{p})^{1/2}V_{c,100}^{9/2}t_{H,15}^{1/2}$ solar masses or, taking the
midpoint of the proposed range $\sigma/m_{p}=14.2cm^{2}/g$, the result
$M_{BH}=2.7\times10^{7}V_{c,100}^{9/2}t_{H,15}^{1/2}M_{\odot},$ with a range
about this $(10^{\pm0.75})$ encouraged by the Spergel Steinhardt analysis.
\ We replace the speed of sound $C_{A}$ with the circular velocity $V_{c}%
^{2}=2C_{A}^{2}$ in the above relation to connect more conveniently to normal
astronomical measurements. \ For both our own galaxy where $V_{c,100}%
\thickapprox2$, and for M87 for which $V_{c,100}\approx7$, the resulting black
hole masses are too large, as compared to observed estimates [4] by a factor
of several hundred. \ This should be treated as a remarkable agreement given
the crudeness both of the analysis and of the estimated collisional DM
properties. \ Accretion in the optically thin limit is only significant for
the lower part of this mass range.

Clearly, the assumption of a single stellar mass black hole in the central
region of the galaxy is an over simplification, since multiple seeds are
expected. \ Given the instability of the initial growth, one BH is likely to
dominate and will eat or eject the others [18]. \ This complication is not
likely to significantly alter the estimates of the final black hole mass.

It is reasonable to ask how the solution changes if we relax the assumption
that the profile is that of a singular isothermal sphere for which the density
profile (eq. 2) is $\rho\varpropto r^{-\alpha}$ with $\alpha=2$. \ The widely
adopted NFW [19] profile for dark matter halos takes $\alpha=1$ in the inner
parts, and other authors find typical values near $\alpha$ \ = 3/2 (see [20,
21] for references). \ One can show [17] that, in the more general case, if
the galactic luminosity scales as $L\varpropto V_{c}^{\mu}$, the mass-to-light
ratio scales as (M/L) $\varpropto V_{c}^{\delta}$, so that the mass scales as
$M\varpropto V_{c}^{\mu+\delta}$, then the final black hole mass will scale
with the galaxy mass as $M_{BH}\varpropto M_{gal}^{K}$ , where
\begin{equation}
K\equiv\frac{9\left[  1-\left(  1-\frac{\alpha}{2}\right)  \left(  \mu
+\delta-2\right)  \right]  }{\left(  \frac{5}{2}\alpha-3\right)  \left(
\mu+\delta\right)  }.\tag*{(6)}%
\end{equation}
Remarkably enough, for the case of greatest interest, where $\mu+\delta=9/2 $,
then K = 1 regardless of the value of $\alpha$. \ Thus, the conclusion reached
in the case of a cusp which is initially that of a singular isothermal sphere
$\left(  \alpha=2\right)  $, that $M_{BH}\varpropto M_{gal}$, the
observational result found by Magorrian et al. [4], is likely to be very close
to that generally obtained for values of the initial cusp parameter departing
moderately from this value.

In the most plausible [20, 21] range for the parameter $\alpha$,
$(\alpha=1.3\pm0.2)$, the core profile would evolve due to the collisional
effects, initially reducing the central density [1, 22] and lowering the rate
of growth of the central BH, \textit{i.e.}, the evolution of the core must be
treated simultaneously with the growth of the central BH. \ Quinlan's [22]
numerical work indicates that the collisional evolution of the core ultimately
approaches the singular core collapse $(\alpha=2.25)$ solution [23] which
would, if it were reached, greatly accelerate the growth of the central black
hole. \ However, for the acceptable range of the parameter $\sigma/m$, this
state would not be reached within a Hubble time for all except the most
extreme systems (see also Burkert [24]). \ Also, the rotation of the dark
matter halo will not be trivial, but, since the viscous and angular momentum
(outward) transport times are the same as the other relevant time scales [13],
this is not likely to provide a significant barrier to accretion. \ Both of
these effects reduce the estimated mass of the BH to a level below that given
by equation [5]. \ Nevertheless, the limit on the observed sizes of central
galactic black holes (perhaps below the Magorian \textit{et. al} [4]
estimates) probably restricts $\sigma/m_{p} $ to the lower end of the range
proposed by Spergel and Steinhardt: \ $\sigma/m_{p}\lesssim1cm^{2}/gm$.
\ However, only detailed numerical calculations (now in progress) will be able
to establish a more precise bound.

It is quite possible that physical processes not included in this elementary
treatment could substantially inhibit the growth of black holes in the
Spergel-Steinhardt scenario, the required values being somewhat smaller than
given by equation [5]. \ We must conclude that the existence of strongly
self-interacting dark matter has the exciting potential for leading to the
growth of central massive black holes in normal galaxies with the observed
scaling parameters. \ 

Three corollary consequences should be noted. \ First, since the hypothesized
dark matter particles do not have radiative interactions, accretion of them
will not produce an electromagnetic luminosity output, $L_{em}$, breaking the
assumed link $\int L_{em}dt=\epsilon_{em}\Delta M/c^{2}$, with normal
estimates of $\epsilon_{em}\approx0.1.$ \ Alternatively phrased, if most of
the accreted matter is dark matter, then a low efficiency $(\epsilon_{em}<<1)$
is to be expected. \ Second, since the mean free path of the particles is
comparable to the system size, the dark matter fluid will be extremely
viscous, with dynamical consequences that may be imagined. \ Finally, dark
matter galactic halos in clusters of galaxies will tend to evaporate due to
heat transfer from the hotter, cluster dark matter. \ Preliminary estimates of
the significance of this effect again limit one to consider the range
$(\sigma/m_{p})\lesssim1cm^{2}/gm$.

\begin{center}
ACKNOWLEDGMENTS
\end{center}

I would like to thank Bohdan Paczynski, Martin Rees, David Spergel, Paul
Steinhardt, Scott Tremaine, Neil Turok and Simon White for very helpful
conversations, the hospitality of the Newton Institute in Cambridge and the
Max Planck Institut for Astrophysics in Garching, where the work was begun,
and NSF grants AST-9424416, AST-9803137, and ASC-9740300.

REFERENCES

[1] \ D.N. Spergel. and P.J. Steinhardt, \textit{Phys. Rev. Letters},
(accepted) Astro-ph/9909386 (1999).

[2\} \ E.D. Carlson, M.E. Machacek and L.J. Hall, \textit{ApJ, }\textbf{398},
43-52 (1992).

[3] \ M.E. Machacek,\ \textit{ApJ, }\textbf{431}, 41-51 (1994).

[4] \ J. Magorrian, S. Tremaine, D. Richstone,R. Bender, G. Bower, A.
Dressler, S.M. Faber, K. Gebhardt, R. Green, C. Grillmair, J. Kormendy, and T.
Lauer, \textit{Astron. J., }\textbf{115}, 2285 (1998).

[5] \ J.P. Ostriker and N. Gnedin,\ \textit{ApJ Letters, }\textbf{472},
L63-L67 (1996).

[6] \ Z. Haiman, M. Rees and A. Loeb, \textit{ApJ}, \textbf{476}, 458 (1997).

[7] \ T.A. Abel, Anninos, M.L. Norman and P., Y. Zhang, \textit{ApJ,
}\textbf{508}, 518-520 (1998).

[8] \ D. Arnett, \textit{Supernovae and Nucleosynthesis} (Princeton University
Press, Princeton, 1996).

[9] \ P. Hodge, \textit{An. Rev. Ast. \& Astroph.}, \textbf{27}, 139 (1989).

[10] \ M. Rees,\ ``Black Holes and Relativity'', in \textit{Astrophysical
Evidence for Black Holes}, edited by R. Wald (University of Chicago Press, 1998).

[11] \ H. Bondi, \textit{Mon. Not. Royal Astron. Soc.}, \textbf{107}, 410 (1947).

[12] \ S. S. Holt and T.R. Kallman, \textit{AIP Conf. Proc. \#431},
\underline{Accretion processes in Astrophysics Systems} (1998).

[13] \ J. Goodman, \underline{Dynamical Relaxation in Stellar Systems},
\textit{Ph.D. Thesis, }Princeton University (1983).

[14] \ J. Stone, J.E. Pringle, M.C. Begelman \textit{MNRAS}, (in press) (2000).

[15] \ D. Lynden-Bell, \textit{Mon. Not. Royal Astron. Soc.}, \textbf{136},
101 (1967).

[16] \ J.N. Bahcall, J. and R.A. Wolf, \textit{ApJ}, \textbf{209}, 214, (1976).

[17] \ J.P. Ostriker and S. Tremaine (in preparation) (1999).

[18[ \ M.C. Begelman, R.D. Blandford and M.J. Rees, \textit{Nature},
\textbf{287}, 309 (1980); G. Xu and J.P. Ostriker, \textit{ApJ}, \textbf{437},
184 (1994).

[19] \ J.F. Navarro, C.S. Frenk and S.D.M. White, \textit{Mon. Not. Royal
Astron. Soc.}, \textbf{275}, 720 (1995).

[20] \ K. Subramanian, R. Cen and J.P. Ostriker, \textit{ApJ}, (accepted)(Astro-ph/9909279)(1999).

[21] \ Y.P. Jing and Y. Suto, Astro-ph/0001288 (2000).

[22] \ G. Quinlan, \textit{New Astronomy}, \textbf{1}, 255 (1996).

[23] \ H. Cohen, \textit{ApJ}, \textbf{242}, 765 (1980).

[24] \ A. Burkert, (Astro-ph/0002409)(2000).
\end{document}